\documentclass[letter]{aa} 
\usepackage{graphicx}
\usepackage{txfonts}
\begin{document}
   \title{Into the central 10 pc of the most distant known radio quasar}

   \subtitle{VLBI imaging observations of J1429+5447 at $z$=6.21}

   \author{S.~Frey\inst{1,5}
           \and
           Z.~Paragi\inst{2,5}
	   \and
	   L.I.~Gurvits\inst{2,3}
 	   \and
	   K.\'E.~Gab\'anyi\inst{1,5}
           \and
	   D.~Cseh\inst{4}
	   }

   \institute{F\"OMI Satellite Geodetic Observatory, PO Box 585, H-1592 Budapest, Hungary\\
              \email{frey@sgo.fomi.hu, gabanyik@sgo.fomi.hu}
         \and
              Joint Institute for VLBI in Europe, Postbus 2, 7990 AA Dwingeloo, The Netherlands\\
              \email{zparagi@jive.nl, lgurvits@jive.nl}
         \and
              Department of Astrodynamics \& Space Missions, Delft University of Technology, Kluyverweg 1, 2629 HS Delft, The Netherlands
         \and
              Laboratoire Astrophysique des Interactions Multi-echelles (UMR 7158),
              CEA/DSM-CNRS-Universit\'e Paris Diderot, CEA Saclay, F-91191 Gif sur Yvette, France\\
              \email{david.cseh@cea.fr}
         \and
              MTA Research Group for Physical Geodesy and Geodynamics, PO Box 91, H-1521 Budapest, Hungary}

   \date{Received May 25, 2011; accepted June 3, 2011}

 
  \abstract
   {There are about 60 quasars known at redshifts $z$$>$5.7 to date. Only three of them are detected in the radio above 1~mJy flux density at 1.4 GHz frequency. Among them, J1429+5447 ($z$=6.21) is the highest-redshift radio quasar known at present. These rare, distant, and powerful objects provide important insight into the activity of the supermassive black holes in the Universe at early cosmological epochs, and on the physical conditions in their environment.}
   {We studied the compact radio structure of J1429+5447 on the milli-arcsecond (mas) angular scale, in order to compare the structural and spectral properties with those of other two $z$$\sim$6 radio-loud quasars, J0836+0054 ($z$=5.77) and J1427+3312 ($z$=6.12).}
   {We performed Very Long Baseline Interferometry (VLBI) imaging observations of J1429+5447 with the European VLBI Network (EVN) at 1.6~GHz on 2010 June 8, and at 5~GHz on 2010 May 27.}
   {Based on its observed radio properties, the compact but somewhat resolved structure on linear scales of $<$100~pc, and the steep spectrum, the quasar J1429+5447 is remarkably similar to J0836+0054 and J1427+3312. To answer the question whether the compact steep-spectrum radio emission is a ``universal'' feature of the most distant radio quasars, it is essential to study more, yet to be discovered radio-loud active galactic nuclei at $z$$>$6.}
   {}


   \maketitle
%

\section{Introduction}

Observations of quasars at the highest redshifts can constrain models of the birth and early cosmological evolution of active galactic nuclei (AGNs) and the growth of their central supermassive (up to $\sim$$10^9$~$M_{\odot}$) black holes. Currently the object CFHQS~J021013$-$045620 holds the redshift record among quasars with $z$=6.44 (Willott et al. \cite{Will10a}). There are about 60 quasars known at redshifts $z$$>$5.7 to date. Despite the steadily growing number of known $z$$\sim$6 quasars (e.g. Fan et al. \cite{Fan01,Fan03,Fan04,Fan06}; Jiang et al. \cite{Jian09}; Willott et al. \cite{Will07,Will09,Will10b}), the measured maximum redshift practically did not increase over the last decade or so (e.g. SDSS~J1148+5251, $z$=6.43; Fan et al. \cite{Fan03}). It remains to be seen whether this is a selection effect due to current limitations of the high-redshift identification techniques, or the first quasars in the Universe indeed started to ``turn on'' at around this cosmological epoch. Intriguingly, many of the intrinsic properties observed in the infrared, optical, and X-ray wavebands make the highest-redshift quasars hardly distinguishable from the lower-redshift ones. However, Jiang et al. (\cite{Jian10}) found two out of twenty-one $z$$\sim$6 quasars which do not show infrared emission originating from hot dust. According to their interpretation, the amount of hot dust may increase in parallel with the growth of the central black hole. Therefore at least some of the most distant quasars known are less evolved than their low-redshift counterparts. Near-infrared spectroscopy of a sample of Canada-France High-z Quasar Survey (CFHQS) objects indicates that these AGNs are accreting close to the Eddington limit and are in the early stage of their evolutionary cycle, building up the mass of the central black hole exponentially from the material of their young host galaxies (Willott et al. \cite{Will10a}). 

Only three of the $z$$>$5.7 quasars (J0836+0054, $z$=5.77, Fan et al. \cite{Fan01}; J1427+3312, $z$=6.12, McGreer et al. \cite{McGr06}; J1429+5447, $z$=6.21, Willott et al. \cite{Will10b}) are found in the radio with 1.4-GHz flux density $S_{1.4}$$>$1~mJy. Most recently, Zeimann et al. (\cite{Zeim11}) detected radio emission ($S_{1.4}$=0.31~mJy) from another distant quasar, J2228+0110 ($z$=5.95). These radio-emitting high-redshift objects are particularly valuable, since the ultimate evidence for AGN jets can be found in the radio by high-resolution Very Long Baseline Interferometry (VLBI) observations. Synchrotron radio emission of the jets originates from the close vicinity of the spinning supermassive black hole. The radio-emitting plasma is fueled from an accretion disk, and is accelerated and collimated by the magnetic field. Two of the $z$$\sim$6 radio sources (J0836+0054, J1427+3312) have already been investigated with VLBI. The radio structure of J0836+0054 on $\sim$10 milli-arcsecond (mas) angular scale (Frey et al. \cite{Frey03,Frey05}) is characterised by a single compact but somewhat resolved component, with steep radio spectrum ($\alpha$=$-0.8$) between the observed frequencies of 1.6 and 5~GHz. (The power-law spectral index $\alpha$ is defined as $S\propto\nu^{\alpha}$, where $S$ is the flux density and $\nu$ the frequency.) The 1.4-GHz and 1.6-GHz VLBI images of the first $z>6$ radio quasar, J1427+3312 (Momjian et al. \cite{Momj08} and Frey et al. \cite{Frey08}, respectively) revealed a prominent double structure. The two resolved components are separated by $\sim$28~mas ($\sim$160~pc). (To calculate linear sizes and luminosities, we assume a flat cosmological model with $H_{\rm{0}}=70$~km~s$^{-1}$~Mpc$^{-1}$,  $\Omega_{\rm m}=0.3$, and $\Omega_{\Lambda}=0.7$.) This structure is similar to that of the compact symmetric objects (CSOs). It is one of the indications of the youthfulness of J1427+3312. The brighter component also detected with VLBI at 5~GHz (Frey et al. \cite{Frey08}) has a steep radio spectrum ($\alpha$=$-0.6$).

A census of VLBI-imaged radio quasars at $z$$>$4.5, and European VLBI Network (EVN) imaging of five new sources at $4.5$$<$$z$$<$$5$ was made recently by Frey et al. (\cite{Frey10}). The slightly resolved mas- and 10-mas-scale radio structures, the measured moderate brightness temperatures ($\sim$10$^8$$-$10$^9$~K), and the steep spectra in the majority of the cases suggest that the sample of compact radio sources at $z$$>$4.5 is dominated by objects that do not resemble blazars, which are characterised by highly Doppler-boosted, compact, flat-spectrum radio emission. At frequencies lower than the turnover value that corresponds to the peak flux density, the rising radio spectrum is determined by synchrotron self-absorption (Slysh \cite{Slys63}). Above the turnover frequency, the emitting plasma becomes optically thin and the spectrum is steep. According to the model of Falcke et al. (\cite{Falc04}), the high-redshift steep-spectrum objects may represent gigahertz-peaked-spectrum (GPS) sources at early cosmological epochs. The first generation of supermassive black holes could have had powerful jets that developed hot spots well inside their forming host galaxy, on linear scales of 0.1$-$10~kpc. Taking the relation between the source size and the turnover frequency observed for GPS sources into account, and for hypothetical sources matching the luminosity and spectral index of ``typical'' quasars in the VLBI sample with $z$$\sim$5 or higher, the angular size of the smallest ($\sim$$100$~pc) of these early radio-jet objects would be in the order of 10~mas, and the observed turnover frequency in their radio spectra would be around 500~MHz in the observer's frame (Falcke et al. \cite{Falc04}).

In this paper, we report on dual-frequency VLBI observations of the currently most distant known radio quasar, one of the only two at $z$$>$6. The object CFHQS~J142952+544717 (\object{J1429+5447} in short) was discovered by Willott et al. (\cite{Will10b}). Its spectroscopic redshift is $z$=6.21 measured from the Ly$\alpha$ emission line. The quasar appears in the Very Large Array (VLA) Faint Images of the Radio Sky at Twenty-centimeters (FIRST) survey (White et al. \cite{Whit97}) list as an unresolved ($<$$5\arcsec$) radio source with an integral flux density of $S_{1.4}$=2.95~mJy. Strong molecular CO (2-1) emission was detected in the host galaxy of J1429+5447 by Wang et al. (\cite{Wang11}), who resolved the source into two components separated by $1\farcs2$. It indicates a gas-rich major merger system. The rapid galaxy formation and starburst activity apparently goes in parallel with the radio-active phase of the accreting supermassive black hole in the western component. The redshift of the source is $z$=6.18 based on the CO (2-1) line.  

Our aim was to reveal its high-resolution radio structure and spectral properties, and then compare them with those of the two other $z$$\sim$6 quasars already known, on linear scales of $\sim$10--100 pc. In the adopted cosmological model, the redshift $z$=6.21 corresponds to 0.875~Gyr after the Big Bang (6.5\% of the present age of the Universe), and 1~mas angular separation is equivalent to a projected linear distance of 5.6~pc.

\section{EVN observations and data reduction}

\begin{figure}[!h]
\centering
  \includegraphics[bb=68 169 522 626, height=80mm, angle=270, clip=]{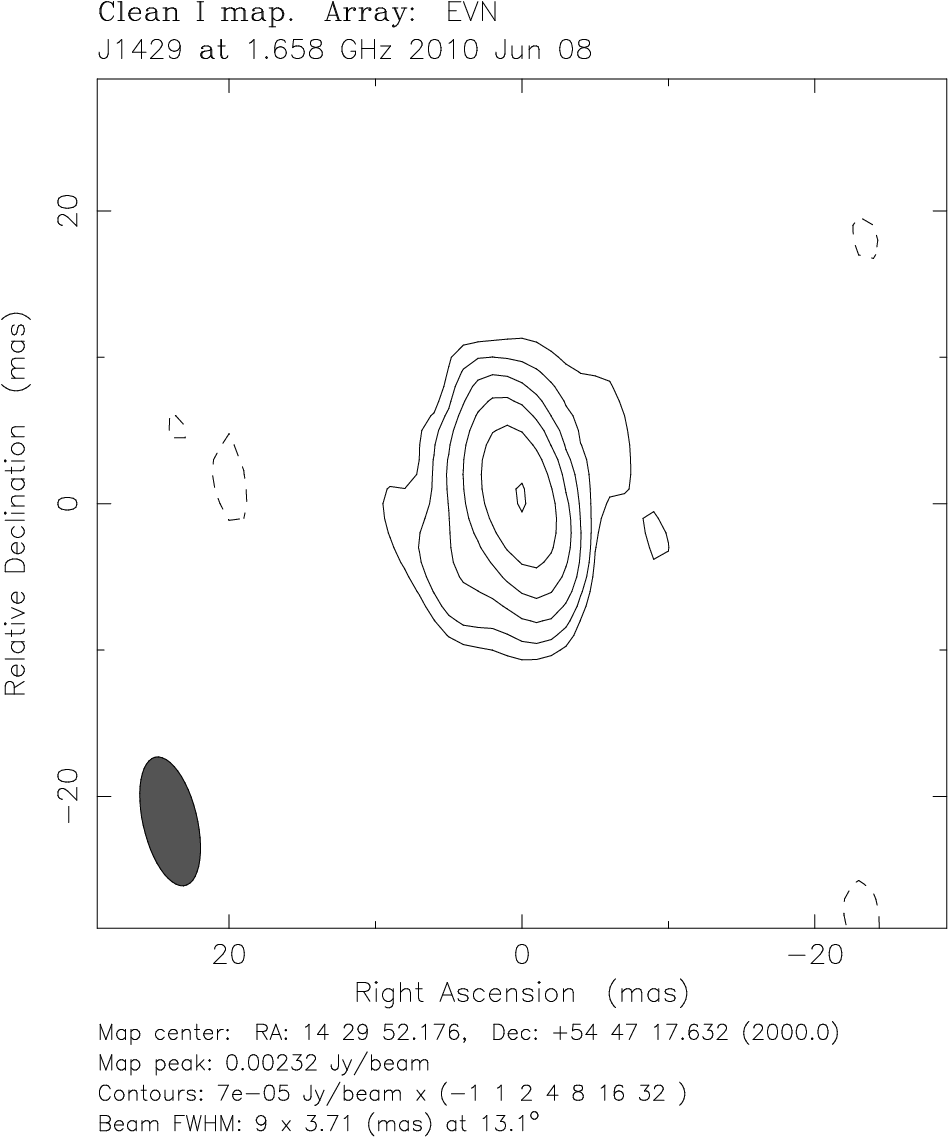}
  \caption{
The naturally weighted 1.6-GHz EVN image of the quasar J1429+5447. The lowest contours are drawn at $\pm70$~$\mu$Jy/beam. The positive contour levels increase by a factor of 2. The peak brightness is 2.32~mJy/beam. The Gaussian restoring beam is 9.0~mas $\times$ 3.7~mas with major axis position angle $13\degr$. The restoring beam (full width at half maximum, FWHM) is indicated with an ellipse in the lower-left corner. The image is centered on the 5-GHz brightness peak.}
  \label{lband}
\end{figure}

\begin{figure}[!h]
\centering
  \includegraphics[bb=68 169 522 626, height=80mm, angle=270, clip=]{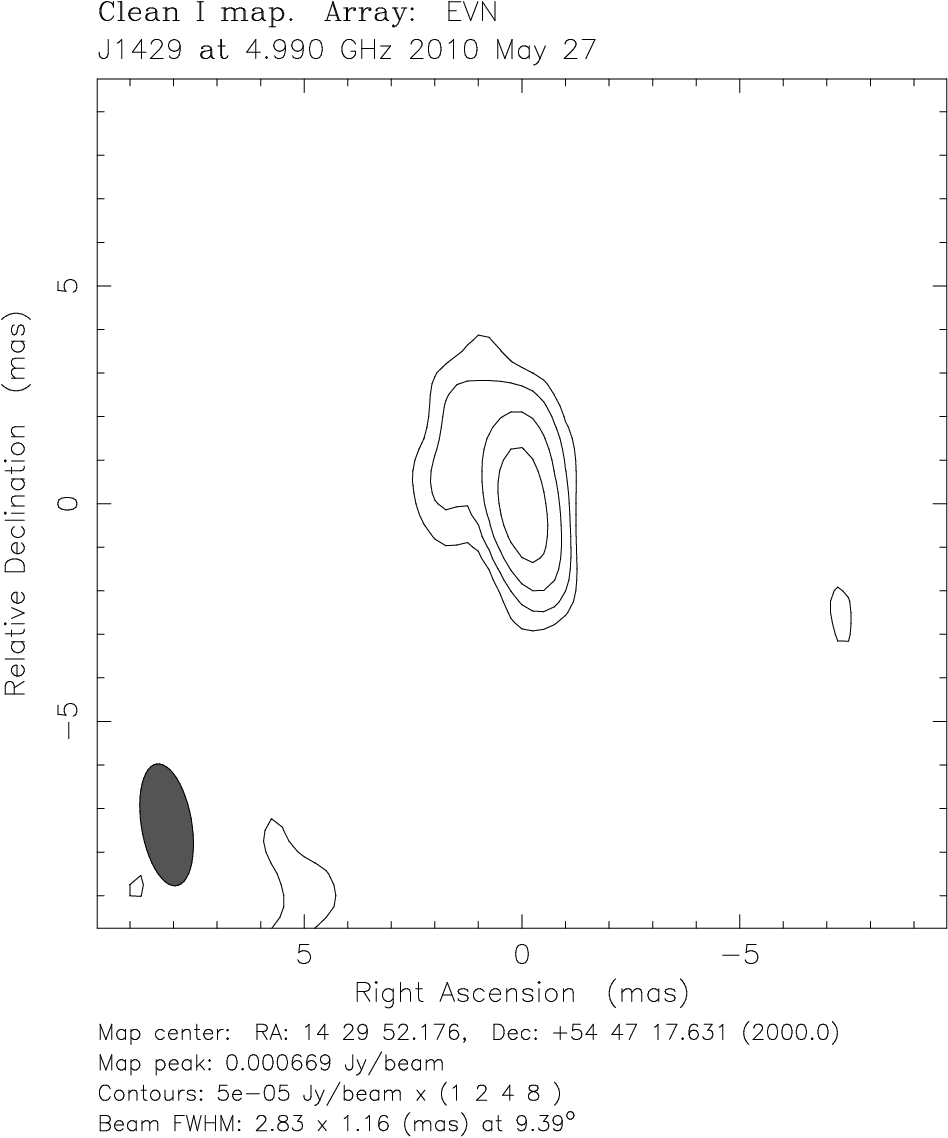}
  \caption{
The naturally weighted 5-GHz EVN image of the quasar J1429+5447. The lowest contours are drawn at $\pm50$~$\mu$Jy/beam. The positive contour levels increase by a factor of 2. The peak brightness is 0.67~mJy/beam. The Gaussian restoring beam is 2.8~mas $\times$ 1.2~mas with major axis position angle $9\degr$. The restoring beam (FWHM) is indicated with an ellipse in the lower-left corner. The image is centered on the brightness peak of which the coordinates are given in the text.}
  \label{cband}
\end{figure}

The EVN observations of J1429+5447 took place on 2010 May 27 (5~GHz frequency), and on 2010 June 8 (1.6~GHz). At a recording rate of 1024~Mbit~s$^{-1}$, eleven antennas of the inter-continental radio telescope network participated in the experiment at 5 GHz: Effelsberg (Germany), Jodrell Bank Lovell \& Mk2 telescopes (UK), Medicina (Italy), Toru\'n (Poland), Onsala (Sweden), Sheshan, Nanshan (P.R. China), Badary, Zelenchukskaya (Russia), and the phased array of the Westerbork Synthesis Radio Telescope (WSRT, The Netherlands). All but the Jodrell Bank Mk2 telescope participated in the 1.6-GHz experiment as well. Both experiments lasted for 6 h. Eight intermediate frequency channels (IFs) were used in both left and right circular polarisations. The total bandwidth was 128~MHz per polarisation. The correlation of the recorded VLBI data took place at the EVN Data Processor at the Joint Institute for VLBI in Europe (JIVE), Dwingeloo, The Netherlands.

The weak target source, J1429+5447 was observed in phase-reference mode to increase the coherent integration time spent on the source and thus to improve the sensitivity of the observations. We refer to Frey et al. (\cite{Frey08}) for the details of the observing and data reduction in a much similar dual-frequency EVN experiment on J1427+3312. The phase-reference calibrator source we used in the present case, J1429+5406 is separated from the target by $0\fdg69$ in the sky. The target--reference cycles of $\sim$5.5~min allowed us to spend $\sim$3.5~min on the target source in each cycle, providing almost 3 h total integration time on J1429+5447. Phase-referencing also allows us to determine the accurate relative position of the target source with respect to the well-known absolute position of the reference source. 

The US National Radio Astronomy Observatory (NRAO) Astronomical Image Processing System (AIPS) was used for the data calibration in a standard way. The calibrated data were then exported to the Caltech Difmap package for imaging (see e.g. Frey et al. \cite{Frey08} for the details and references). The naturally-weighted images at 1.6~GHz (Fig.~\ref{lband}) and 5~GHz (Fig.~\ref{cband}) were made after several cycles of CLEANing in Difmap. No self-calibration was applied. The lowest contours indicate $\sim$$3\sigma$ image noise levels. The theoretical thermal noise values were $\sim$10~$\mu $Jy/beam ($1\sigma$) at both frequencies.

\section{Results and discussion}

There is a single dominant radio feature detected in the quasar J1429+5447 at both 1.6 and 5~GHz. The images in Fig.~\ref{lband}-\ref{cband} are centered on the 5-GHz brightness peak of which the phase-referenced absolute equatorial coordinates are right ascension $14^{\rm h}29^{\rm m}52\fs17629$ and declination $54\degr47\arcmin17\farcs6309$ (J2000), each with the accuracy of 0.4~mas, determined by the phase-reference calibrator source position accuracy, the target--calibrator separation, the angular resolution of the interferometer array, and the signal-to-noise ratio. The images show a slightly resolved mas-scale structure. Difmap was used to fit circular Gaussian brightness distribution model components to the interferometric visibility data at both frequencies. The 5-GHz data are well described by a component with 0.99~mJy flux density and 0.67~mas diameter (Table~\ref{modelfit}). These imply the rest-frame brightness temperature $T_{\rm B}$=(7.7$\pm$0.7) $\times$ $10^{8}$~K. It confirms the AGN origin of the quasar's radio emission since the brightness temperatures for normal galaxies don't exceed $\sim$$10^{5}$~K (Condon \cite{Cond92}). At 1.6~GHz, the best-fit model is composed of two circular Gaussians (Table~\ref{modelfit}). A hint on a corresponding weak extension to the south-east can also be found in the image (Fig.~\ref{lband}), where, however, the shape of the lowest (3$\sigma$) contour line should not be over-interpreted. Considering the calibration uncertainties, the sum of the flux densities in the VLBI components (3.30~mJy) is consistent with or somewhat higher than the FIRST value (2.95$\pm$0.15~mJy). We effectively detect the entire radio emission in the 1.4--1.6~GHz band from a $\sim$10-mas region of J1429+5447, corresponding to the linear size less than 60~pc. At 5~GHz, the comparison between the VLBI component flux density (0.99~mJy) and the result from our analysis of the WSRT array data taken during our EVN experiment (1.2~mJy) allows us to draw a similar conclusion: all the radio emission originates from a $\sim$10-pc region in the quasar's centre. In the case of J1429+5447, any other possible secondary component is excluded within an angular radius of $\sim$$2\arcsec$, at the brightness level of $\sim$90~$\mu$Jy/beam or higher (assuming at least 5$\sigma$ detection and 10\% coherence loss) in the 1.6-GHz VLBI image. In particular, we did not detect compact radio emission at the location of the eastern CO (2-1) line-emitting component (Wang et al. \cite{Wang11}). On the other hand, the brightness peaks in our VLBI images coincide with their western component, and also with the 32-GHz continuum emission peak. Wang et al. (\cite{Wang11}) measured 257$\pm$15~$\mu$Jy for the flux density at 32~GHz in the continuum source which is unresolved with the Expanded VLA (C configuration, $0\farcs71$$\times$$0\farcs67$ synthesized beam).      

Based on our VLBI measurements, the two-point spectral index for the dominant component of the source is $\alpha$=$-1.0$. The 32-GHz continuum flux density (Wang et al. \cite{Wang11}) is consistent with the steep synchrotron radio spectrum. The total rest-frame 5-GHz monochromatic luminosity of J1429+5447 is $4.5 \times 10^{26}$~W~Hz$^{-1}$, comparable to other high-redshift sources (e.g. Frey et al. \cite{Frey10}).  

\begin{table}
  \caption[]{The fitted VLBI model parameters for J1429+5447.}
  \label{modelfit}
\begin{tabular}{ccccc}        
\hline\hline                 
Flux density       & \multicolumn{2}{c}{Relative position } & Size          & $T_{\rm B}$  \\
(mJy)              & north (mas) & east (mas)               & (mas)         & ($10^{8}$~K) \\
\hline                       
3.03$\pm$0.05      &  ...               &  ...              & 2.63$\pm$0.03 & 14.0$\pm$0.6 \\
0.27$\pm$0.04      & $-$4.74$\pm$0.06   &  4.26$\pm$0.06    & 1.29$\pm$0.12 &  5.2$\pm$1.7 \\
\hline                       
0.99$\pm$0.06      &  ...               &  ...              & 0.67$\pm$0.01 &  7.7$\pm$0.7 \\
\hline   
\end{tabular}
\\
\tablefoot{
The parameters are derived from circular Gaussian model-fitting to VLBI visibility data in Difmap, at 1.6~GHz (top) and 5~GHz (bottom). The statistical errors are estimated according to Fomalont (\cite{Foma99}). Additional flux density calibration uncertainties are assumed as 5\%.
Col.~1 -- model component flux density (mJy); 
Col.~2-3 -- separation from the main component to north and east (mas);
Col.~4 -- diameter (mas, FWHM);
Col.~5 -- brightness temperature ($10^{8}$~K).
}
\end{table}

The inferred $T_{\rm B}$$\simeq$$10^{9}$~K brightness temperature is substantially smaller than the equipartition value estimated for relativistic compact jets ($T_{\rm B,eq}$$\simeq$5 $\times$ $10^{10}$~K; Readhead \cite{Read94}). This suggests that relativistic beaming does not play a major role in the appearance of the source. According to the orientation-based unified picture of radio-load AGNs (Urry \& Padovani \cite{Urry95}), the radiation from the jets that are inclined with a sufficiently large angle ($\theta$) to the line of sight is deamplified with a Doppler factor $\delta$$<$1. The viewing angle and the Doppler factor are related to the bulk Lorentz factor ($\Gamma$) of the plasma as

\begin{equation}
\delta  = \frac{1}{\Gamma(1-\beta \cos \theta)},    
\end{equation}
\noindent
where $\beta<1$ is the bulk speed of the material in the jet, expressed in the units of the speed of light $c$. 
For J1429+5447, assuming energy equipartition between the particles and the magnetic field in the radio-emitting region, our VLBI measurements imply $\delta=T_{\rm B}/T_{\rm B,eq}\simeq0.02$. This, in the extreme case of the jets in the plane of the sky ($\theta$=$90\degr$), would require a minimum Lorentz factor $\Gamma$$\ga$50. Values up to $\Gamma$=$50$ are indeed inferred from jet kinematics in a sample of powerful lower-redshift radio AGNs (Lister et al. \cite{List09}), but such an extreme Lorentz factor and the special geometry with $\theta$=$90\degr$ make this scenario very unlikely for J1429+5447. Since the other two $z$$\sim$6 radio quasars have similar or even smaller brightness temperatures with respect to the equipartition value (Frey et al. \cite{Frey05,Frey08}), the above reasoning is valid for them as well. It is also the case for a significant portion of other radio AGNs at $z$$>$4.5 studied with VLBI so far (Frey et al. \cite{Frey10}). This, and the steep spectrum are arguments against the possibility that we see intrinsically powerful but Doppler-deboosted (highly inclined) relativistic radio jets in these sources. Although currently there are about 60 quasars known at $z$$\ga$6, we don't know any of them with Doppler-boosted radio emission. The steep spectrum between the observed 1.6 and 5~GHz frequencies (12 and 36~GHz in the rest frame of the quasar J1429+5447) is consistent with the assumption that we see the compact ``hot spots'' confined within a region of $<100$~pc in a young GPS source at an early cosmological epoch. The spectral peak frequency is possibly redshifted to the $\sim$100~MHz range in the observer's frame (Falcke et al. \cite{Falc04}). This could be verified with low-frequency radio interferometric measurements.

\section{Conclusions}

Based on our EVN observations at 1.6 and 5~GHz, the mas-scale radio structure of the highest-redshift radio quasar known to date, J1429+5447 ($z$=6.21), is quite similar to what we have seen in the other two $z$$\sim$6 quasars, J0836+0054 and J1427+3312. The source is somewhat resolved on linear scales of $<$100~pc, although the total radio emission is confined to this central region. The two-point spectral index $\alpha$=$-1.0$ indicates a steep radio spectrum. The measured brightness temperature shows that relativistic beaming does not influence the appearance of this quasar. Interestingly, the compact steep-spectrum radio emission is common in all the three $z$$\sim$6 quasars studied with VLBI to date. The puzzle whether it's a rule or an exception could be solved using a prospective larger sample of extremely distant radio-loud AGNs in the future.

\begin{acknowledgements}
The EVN is a joint facility of European, Chinese, South African, and other radio astronomy institutes funded by their national research councils. This work was supported by the European Community's Seventh Framework Programme, Advanced Radio Astronomy in Europe, grant agreement no.\ 227290, the European Community's Seventh Framework Programme (FP7/2007-2013) under grant agreement no.\ ITN 215212 ``Black Hole Universe'', and the Hungarian Scientific Research Fund (OTKA, grant no.\ K72515). 
\end{acknowledgements}

\end{document}